\documentclass[twocolumn,showpacs,showkeys,amsmath,amssymb]{revtex4}

\begin{document}

\title{Baryon magnetic moments  in the SU(3) and the SU(2)$\times$U(1)
   flavor groups}

\author{J. G. Contreras, R. Huerta, L. R. Quintero}
\affiliation{
Departamento de F\'{\i}sica Aplicada\\
Centro de Investigaci\'on y Estudios Avanzados del Instituto
Polit\'ecnico Nacional, Unidad M\'erida\\ Apartado Postal 73 Cordemex\\
97310 M\'erida, Yucat\'an, M\'exico
}

\date{\today}

\begin{abstract}
  Working within the non relativistic quark model a two parameter fit
  to the magnetic moments of baryons is presented.  The fit has an
  excellent $\chi^2$.  The model is based on taking different flavor
  groups to describe the different magnetic moments.  The selection of
  which group to assign to each baryon is guided by the structure of
  its wavefunction. The model corresponds to assigning different
  effective masses to a quark depending on which baryon is being
  considered.  Using the values extracted from the fit, the magnetic
  moments of the $\Omega^-$ and the $\Delta^{++}$ have been
  predicted and the comparison to the existing experimental values is
  quite satisfactory.
\end{abstract}
\maketitle

\section{Introduction}
There has recently been a renewed interest in the magnetic moments and
spin structure of baryons within a variety of models. For example, the
chiral quark model~\cite{franklin0,linde}, quenched lattice gauge
theory~\cite{leinweber}, the 1/N$_\mathrm{c}$ expansion~\cite{jenkins}
and extensions to the non-relativistic quark model (NRQM)~\cite{ha}
to name a few.  These models are more ambitious than the NRQM.
Nonetheless it has been argued that due to some subtle cancellations
the NRQM is a good approximation to the magnetic moments~\cite{cheng},
so a simple model may extract the physics of the problem
more easily than a complicated one.

It is well known since a long time that the magnetic moments of the
octet baryons can be described only approximately via a SU(3) flavor
group~\cite{coleman}.  Using this approach within the NRQM, it is
assumed that the breaking of the flavor symmetry acts equally in all
states of the octet. However this is not physically acceptable. Take
for example the relation $\mu_p/\mu_n=-1.5$. It can be obtained either
from the SU(2) as well as from the SU(3) flavor group. It is known
that the SU(3) symmetry is valid at the 30\% level, so the relation
should be valid at this level. But the relation is experimentally
valid at the 1.5\% percent level as expected from the SU(2)
symmetry. Therefore, in this case, it can not be accepted that the
breaking of SU(3) acts as it does for the case of, say, the magnetic
moment of the $\Lambda$.

Accordingly, in this letter it is proposed to change the idea to use
one broken flavor group to describe all magnetic moments, to the idea
of describing some magnetic moments with one exact flavor group and
some with another exact flavor group. Furthermore we give criteria,
based on the structure of the wavefunctions, as to which flavor group
to use to describe the magnetic moment of a given baryon.  A physical
interpretation of the model in terms of effective quark masses is
provided.

Section \ref{s:wf} introduces the wave functions of the baryons under
the SU(2)$\times$U(1) flavor group (times the SU(2) spin group).
Section \ref{s:mm} presents the magnetic moments of the baryons
keeping the masses of the three quarks as different
parameters. Remarkably, although the wave functions are different
under the SU(3) and the SU(2)$\times$U(1) flavor groups, the magnetic
moments turn out to be the same.

In section \ref{s:fit} the fits to the measured magnetic moments are
presented. First we note that each magnetic moment can be written
either in the form $\mu=a\mu_d(1+b(\mu_s/\mu_d))$ or as
$\mu=c\mu_s(1+d(1-\mu_s/\mu_d))$ (from here on the approximation
$m_u=m_d$ will be assumed). It is found that each baryon falls in only
one of the two following sets: $i$) The baryon has either a small
contribution to its magnetic moment coming from the $\mu_s/\mu_d$
factor, i.e. the coefficient $b$ is less than one or $ii$) it has a
small contribution to its magnetic moment coming from the
$1-\mu_s/\mu_d$ factor, i.e. the coefficient $d$ is less than
one. Note that the exact SU(3) flavor limit implies $\mu\rightarrow
c\mu_s$ ($d=0$), and the SU(2)$\times$U(1) exact flavor limit with
$m_d<<m_s$ implies $\mu\rightarrow a\mu_d$ ($b=0$).  This two set
division of the magnetic moments can be readily explained using the
wave functions of the baryons and it provides criteria to know which
group should be used to calculate the magnetic moment of a given
baryon.  Then a fit is performed using the formulas for the magnetic
moments from the SU(2)$\times$U(1) exact flavor limit, with
$m_d<<m_s$, for the baryons in the first set, and those of the exact
SU(3) flavor group for the baryons in the other set.

It was found that using a two parameter fit to the experimental data a
better agreement in terms of $\chi^2$ was obtained, than other two
parameter fits in the literature and a comparable agreement to fits
requiring four parameters (see for example~\cite{sehgal,pondrom}).
Using the values of the parameters obtained from the fit, the magnetic
moments of the $\Omega^-$ and the $\Delta^{++}$ have been
predicted. The comparison to the existing experimental values is quite
satisfactory.

Section \ref{s:dis} is devoted to a discussion of our results. This
letter is closed with a brief summary and an outlook of future work in
section \ref{s:con}.

The current status on the experimental side is as follows. Seven of
the magnetic moments are measured with around 1\% accuracy or
better~\cite{pdg}. The transition magnetic moment for $\Sigma^0
\rightarrow \Lambda$ is known to a 5\% precision~\cite{petersen}. The
$\Omega^-$ was measured some time ago~\cite{diehl} and recently a new
measurement has been presented~\cite{wallence}.  Finally the magnetic
moment of the $\Delta^{++}$ has also been measured~\cite{bosshard}.

\section{\label{s:wf} Wave functions of baryons}

The procedure to obtain the wave functions for a given flavor group is
standard and can be found in many textbooks. It is well known that the
SU(3) flavor group (times the SU(2) spin group) produces, in the case
of baryons, an {\bf 8} and a {\bf 10} multiplet.  On the other hand
the SU(2)$\times$U(1) flavor group (times the SU(2) spin group) yields
different wave functions (see the appendix) which form the following
multiplets: {\bf 2} ($N$), {\bf 4} ($\Delta$), {\bf 3} ($\Sigma$),
{\bf 1} ($\Lambda$), {\bf 2} ($\Xi$), {\bf 3} ($\Sigma^*$), {\bf 2}
($\Xi^*$), {\bf 1} ($\Omega$).

\section{\label{s:mm}The magnetic moments of baryons}

The expectation value for the magnetic moment $\mu$ of a baryon $B$ in
the S wave is given by the expression $\mu_{B} = <\psi_{B}|
\sum_{i=1}^3 \hat{\mu_{q}}(i) \hat{\sigma_{z}}(i)| \psi_{B}>$ where
$\hat{\mu_{q}}$ is the operator for the magnetic moment of the quarks,
$\hat{\sigma_{z}}(i)$ is Pauli's spin operator and $i$ runs over
$\{u,d,s\}$.  The wave functions for the SU(2)$\times$U(1) group are
different to those found with the SU(3) flavor group (see for
example~\cite{fay}). In spite of this fact, the magnetic moments of
the baryons are the same irrespectively of which set of wave functions
are used to calculate them.

The formulas for the magnetic moments are given in
table~\ref{tab:corr}.  Each magnetic moment can be written using the
functional form $\mu=a\mu_d(1+b(\mu_s/\mu_d))$ and as
$\mu=c\mu_s(1+d(1-\mu_s/\mu_d))$.  Each baryon falls in only one of
the two following sets:
\begin{itemize}
\item[{\bf A}] The baryon has a contribution to its magnetic moment
coming from the $\mu_s/\mu_d$ factor with $b<1$. The baryons in this
group are $p$, $n$, $\Sigma^+$, $\Sigma^-$, $\Sigma^0$,
$\Sigma^0\rightarrow\lambda^0$, $\Delta^{++}$, $\Delta^+$, $\Delta^0$,
$\Delta^-$, $\Sigma^{*+}$, $\Sigma^{*o}$ and $\Sigma^{*-}$.
\item[{\bf B}] The baryon has a contribution to its magnetic moment
coming from the $1-\mu_s/\mu_d$ factor with $d<1$. The baryons in this
group are $\Lambda$, $\Xi^0$, $\Xi^-$, $\Xi^{*0}$, $\Xi^{*-}$ and
$\Omega^-$.
\end{itemize}

This division of the baryons in two sets has an explanation in
terms of wave functions.  The SU(2)$\times$U(1) exact flavor limit with
$m_d<<m_s$ implies $\mu\rightarrow a\mu_d$ so this case can be naturally
identified with set {\bf A}. Note that none of the wave functions in
this case has a dominance of the strange quark. The behavior
corresponds then to a decoupling of $s$ which is described by the
SU(2)$\times$U(1) exact flavor limit with $m_d<<m_s$. On the other
hand the exact SU(3) flavor limit implies $\mu\rightarrow c\mu_s$, so it
can be identified with the set {\bf B}. Here the $s$ quark dominates
the wave function ($\Xi^0$, $\Xi^-$) or the isospin structure of the
wave function cancel the influence of the light quarks when
calculating the magnetic moments ($\Lambda$).
Note that each quark mass $m_q$ is a parameter of the group. So in
principle a more precise notation would be in the lines $m_q^{SU(3)}$
and so on. This notation is rather cumbersome, so the same symbol has
been used for both groups. This does not mean that the value of $m_q$
must be the same in both groups, so there is no contradiction in
having $\mu_s/\mu_d$ small for one group and $1-\mu_s/\mu_d$ small for
the other.

\section{Fit to the magnetic moments of baryons}
\label{s:fit}

There is an important technical point, before doing the fit. The
magnetic moments of both the proton and the neutron have a very small
experimental error. This precision of more than one part per million
is huge when compared to the accuracy of the isospin symmetry of the
 ($p$,$n$) doublet. This turns meaningless a $\chi^2$ approach to the
fit. To avoid this problem, it was proposed in~\cite{franklin} to add
in quadrature a common absolute error to all the moments. Following
this lead (see also~\cite{pondrom}) an absolute error of
$\sigma=0.03\mu_N$ has been added in quadrature to the real
experimental error. The measured values of the magnetic moments of
baryons are shown in table~\ref{tab:meas}

First a three parameter fit was performed. For the elements of set
{\bf A} the form $\mu=a\mu_d(1+\epsilon b(\mu_s/\mu_d))$ was used. For
the elements of set {\bf B} the form $\mu=c\mu_s(1+\epsilon
d(1-\mu_s/\mu_d))$.  $a$, $b$, $c$ and $d$ can be easily read off
table~\ref{tab:corr}. The three parameters are then $\mu_d$, $\mu_s$
and $\epsilon$. The parameter $\epsilon$ turned out to be compatible
with zero ($\epsilon$=0.028$\pm$0.098) while
the values for $\mu_d$ and $\mu_s$ remained exactly as in the two
group fit shown below. This experimental evidence strengths
our assumption of separating the magnetic moments in two different
nonoverlapping sets.  Thus a new 2 parameter fit was performed using
$\mu=a\mu_d$ (set {\bf A}) and $\mu=c\mu_s$ (set {\bf B}).

This two parameter fit can be viewed as two independent one parameter
fits. For the case of the 5 magnetic moments in set {\bf A} a $\chi^2$
per degree of freedom (dof) of 0.42 was found. The other 3 magnetic
moments in set {\bf B} yield $\chi^2/$dof=1.9. To be able to compare
the quality of the fit for this model with other results in the
literature which quote a single value for $\chi^2$, both fits have been
performed simulaneously. In this case $\chi^2/$dof=1.4. The fitted
values of the parameters are $\mu_d=-0.930\pm0.007$ and
$\mu_s=-0.628\pm0.013$.  The values obtained for the magnetic moments
using these parameters are shown in table~\ref{tab:meas} under the
heading $\mu_\mathrm{2G}$. The subscript is meant to stress the fact
that the fit was performed using simultaneously two flavor groups;
SU(2)$\times$U(1) identified with the elements of set {\bf A} and
SU(3) corresponding to the elements of set {\bf B}.  The errors shown
are the maximum spread in the values of the magnetic moments obtained
by varying the parameters within their errors.

\section{\label{s:dis}Discussion}

{\bf 1.} To be able to do the fit, an extra error of
$\sigma=0.03\mu_N$ has been added in quadrature to the experimental
error. This value makes sense as much as in the size of accuracy of
considering the proton and the neutron as an isospin doublet, as in
comparison to the errors of the other measured magnetic moments. Nonetheless
to study the sensitivity of the results to this error, its value was
changed to 0.02 and 0.04 $\mu_N$. As expected, the main effect was in
the $\chi^2/$dof which changed from 1.4 to 2.6 and 0.9 respectively.
The value of the parameters remained the same and their errors varied
from $\pm$0.93 to $\pm$0.17 for $\mu_d$ and $\pm$0.005 to
$\pm$0.009 for $\mu_s$. This shows that the fit is quite stable under
variation of this assumption.  It must be noted that other analysis
have used this extra error up to $\sigma=0.1\mu_N$ to equalize the
weights, within the fit, of the different magnetic moments and to {\em
force} a $\chi^2$/dof of the order of one~\cite{sehgal,karl}.

{\bf 2.} The wave functions obtained using the SU(2)$\times$U(1)
flavor group are different to those from the SU(3) flavor
group. Nevertheless the magnetic moments in both approaches turn out to be
the same when considering the three quark masses as different
parameters. Note that when taking into account the physical hierarchy
of quark masses the magnetic moments of both approches differed and
could be classified in two different sets.

{\bf 3.} From this analysis it is clear that different baryons can be
associated with different flavor groups. This new idea differs from
the traditional method of fixing one flavor group for all baryons and
then breaking it. This result is strengthened by the results of the
the fit with the parameter $\epsilon$. One could argue that the
variation of, say, the coefficients $b$ from baryon to baryon can be
big (for example there is a factor of 2 between $b$ for $\Sigma^+$ and
$b$ for $\Sigma^-$) and that it is too much to ask $b$ and $d$ to be
zero in all cases. Nontheless the fit including the parameter
$\epsilon$ implies exactly that, and from this it follwos naturally
the separation of the baryons in two groups governed by different
exact flavor symmetries.

{\bf 4.} A criteria to decide which flavor group to use for
calculating the magnetic moment of a given
baryon is provided. If the $s$ quark dominates the wave function SU(3)
is a good choice, if not, then SU(2)$\times$U(1) is a better flavor
group. Using these guidelines the full sets are {\bf A}=\{$N$, $\Sigma$,
$\Delta$, $\Sigma^*$\}, {\bf B}=\{$\Lambda$, $\Xi$, $\Xi^*$, $\Omega$\}.

{\bf 5.} Physically this means that the effective masses of the quarks
in a baryon depend on which other quarks are bound to them to form the
baryon.  Pictorically, the quarks dress themselves depending on the
company.  This idea is not so strange as it sounds and it has already
been explored \cite{das,lindej}. In the NRQM the quarks are in a
potential with an energy (the total mass) which changes from baryon to
baryon, so it is natural that the {\em effective} quark masses may
depend on the baryon. There are even experimental evidence that quarks
affect and are affected by their surroundings, i.e., the measurement
of the light quark sea asymmetry \cite{asy}. From the values for the
magnetic moments the quark masses can be computed for each group. It
is found that in the case of exact SU(3) flavor symmetry the three
masses are 498 MeV. For SU(2)$\times$U(1) $m_u=m_d=$336 MeV. In our model
this means that when the $s$ quark dominates the wave function the
effective $u$ and $d$ quarks are heavier than in the absence of the
strange quark. In other words, the presence of a heavier quark induces
an increase on the effective binding energy assinged to the lighter
quarks.

{\bf 6.} The two flavor group model is based on the phenomenological
 idea that the binding energy of quarks, which is effectively assigned
 to their masses in NRQMs, depends on the surronding media.  This
 approach is validated by the excellence of the fit and accuracy of
 its predictions. The magnetic moments of the $\Delta^{++}$ and the
 $\Omega$ have been compared to experimental measurements which have
 not been used in the fits, i.e they are independent and can be used
 to test the model.  It is predicted that
 $\mu_\mathrm{2G}(\Delta^{++})=5.58\pm0.04$ and
 $\mu_\mathrm{2G}(\Omega)=-1.88\pm0.04$ in very good agreement with
 the measured values of 4.52$\pm$0.95 and -2.02$\pm$0.06 respectively.
 
\section{\label{s:con}Conclusions}

The wave functions of baryons for the SU(2)$\times$U(1) flavor group
have been presented. From the wave funtions the magnetic moments of
the baryons have been calculated.  A two parameter fit to the magnetic
moments of the baryons has been performed. The new idea behind the fit
is to use two flavor groups to describe the magnetic moments. A
criteria to assign a given baryon to a flavor group, based in the
structure of its wave function, has been provided.  In terms of
$\chi^2$ the 2 parameter model presented here has an accuracy of the
same order than other 4 parameter fits in the literature. The
parameters have been used to predict the magnetic moments of the
$\Delta^{++}$ and the $\Omega$. An excellent agreement with the
measured values has been found. Given the different multiplet
structure of the two flavor groups used and the different wave
functions they provided, this approach could be applied to the
description of other phenomena like, for example, semileptonic decays
or fragmentation function of baryons.Furthermore, in view of the
success of the model, the idea of two different exact flavor groups
may be used as a guide to simplify calculations and define
approximations in other more formal approaches based on first
principles calculations within QCD.

\begin{table*}
\caption{Expressions for magnetic moments $\mu$ of baryons. The
standard form corresponds to SU(3) with all masses different. The
following columns assume $m_u=m_d$. The formula for SU(2)$\times$U(1) are on
the limit $m_d<<m_s$. In the case SU(3) all masses are equal. The
expressions used in the fit with 3 parameters are shown in the last
column. }
\begin{ruledtabular}
\begin{tabular}{ccccc}
Baryon & Standard Form & SU(2)$\times$U(1) & SU(3) & $\epsilon$ fit \\
\hline
$p$ & $\frac{1}{3}(4\mu_u-\mu_d)$
    &$-3\mu_d$ 
    & $-3\mu_s$ 
    & $-3\mu_d$ \\
$n$ & $\frac{1}{3}(4\mu_d-\mu_u)$
    & $2\mu_d$ 
    & $2\mu_s$ 
    & $2\mu_d$ \\
$\Lambda$ &$\mu_s$
    &0
    & $\mu_s$ 
    & $\mu_s$\\
$\Sigma^+$& $\frac{1}{3}(4\mu_u-\mu_s)$
    &$-8/3\mu_d$ 
    & $-3\mu_s$ 
    & $-8/3\mu_d(1+\frac{\mu_s}{8\mu_d}\epsilon)$ \\
$\Sigma^-$& $\frac{1}{3}(4\mu_d-\mu_s)$
    &$4/3\mu_d$
    & $\mu_s$ 
    &$4/3\mu_d(1-\frac{\mu_s}{4\mu_d}\epsilon)$ \\
$\Sigma^0$&$(2\mu_u+2\mu_d-\mu_2)/3$ 
    & $-2/3\mu_d$ 
    & $\mu_s$ 
    &  \\
$\Xi^0$& $\frac{1}{3}(4\mu_s-\mu_u)$
    &$2/3\mu_d$
    &$2\mu_s$
    & $2 \mu_s[1-\frac{1}{3}(1-\frac{\mu_d}{\mu_s})\epsilon]$  \\
$\Xi^-$&$\frac{1}{3}(4\mu_s-\mu_d)$
    &$-1/3\mu_d$
    &$\mu_s$
    & $\mu_s[1+\frac{1}{3}(1-\frac{\mu_d}{\mu_s})\epsilon]$ \\
$\Sigma^0\rightarrow\Lambda$& $\frac{1}{\sqrt{3}}(\mu_d-\mu_u)$
    &$\sqrt{3}\mu_d$
    &$\sqrt{3}\mu_s$
    &$\sqrt{3}\mu_d$  \\
$\Delta^{++}$&$3\mu_u$
    &$-6\mu_d$
    &$-6\mu_s$
    & \\
$\Delta^{+}$&$2\mu_u+\mu_d$
    &$3\mu_d$ 
    &$3\mu_s$
    & \\
$\Delta^{0}$&$\mu_u+2\mu_d$
    &0
    &0
    &\\
$\Delta^{-}$&$3\mu_d$
    &$3\mu_d$
    &$3\mu_s$
    & \\
$\Sigma^{*+}$&$2\mu_u+\mu_s$
    &$-4\mu_d$
    &$-3\mu_s$
    & \\
$\Sigma^{*0}$&$\mu_u+\mu_d+\mu_s$
    &$-\mu_d$
    &0
    & \\
$\Sigma^{*-}$&$2\mu_d+\mu_s$
    &$2\mu_d$
    &$3\mu_s$
    & \\
$\Xi^{*0}$&$2\mu_s+\mu_u$
    &$-\mu_d$
    &0
    & \\
$\Xi^{*-}$&$2\mu_s+\mu_d$
    &$\mu_d$
    &$3\mu_s$
    & \\
$\Omega$ & $3\mu_s$
    &0
    &$3\mu_s$
    &\\
\end{tabular}
\end{ruledtabular}
\label{tab:corr}
\end{table*}

\begin{table}
\caption{Measured values for baryon magnetic moments in units of
  $\mu_N$, along with the prediction of our two group model with two
  parameters.The experimental values above the middle line were used
  in the fits. The values below the middle line are parameter free
  predictions of our model}
\begin{ruledtabular}
\begin{tabular}{ccc}
Baryon & $\mu_\mathrm{exp}$ & $\mu_\mathrm{2G} $  \\
\hline
$p$&2.79$\pm$6.3x10$^{-8}$&2.79$\pm$0.02\\
$n$&-1.91$\pm$4.5x10$^{-7}$&-1.86$\pm$0.01\\
$\Lambda$&-0.613$\pm$0.004&-0.63$\pm$0.01\\
$\Sigma^+$&2.46$\pm$0.01&2.48$\pm$0.02\\
$\Sigma^-$&-1.16$\pm$0.025&-1.24$\pm$0.01\\
$\Xi^0$&-1.25$\pm$0.014&-1.26$\pm$0.02\\
$\Xi^-$&-0.651$\pm$0.0025&-0.63$\pm$0.01\\
$\Sigma^0\rightarrow\Lambda$&-1.61$\pm$0.08&-1.61$\pm$0.01\\
\hline
$\Sigma^0$&&0.620$\pm$0.005\\
$\Delta^{++}$ &4.52$\pm$0.95& 5.58$\pm$0.04 \\
$\Delta^+$ && 2.79$\pm$0.02 \\
$\Delta^0$ && 0 \\
$\Delta-$ && -2.79$\pm$0.02 \\
$\Sigma^{*+}$ &&3.72 $\pm$0.03 \\
$\Sigma^{*0}$ && 0.93$\pm$0.01 \\
$\Sigma^{*-}$ && -1.86$\pm$0.01 \\
$\Xi^{*0}$ && 0 \\
$\Xi^{*-}$ && -1.88$\pm$0.04 \\
$\Omega$ &-2.02$\pm$0.06& -1.88$\pm$0.04 \\
\end{tabular}
\end{ruledtabular}
\label{tab:meas}
\end{table}

\begin{acknowledgments}
This work has been partially supported by CONACyT
\end{acknowledgments}

\appendix*

\section{\label{a:wf}
Wave functions for the SU(2)$\times$U(1) flavor group}

The  baryon wave functions are given by: 

$\psi_{p\uparrow} = \frac{1}{3\sqrt{2}}[uud
(2\uparrow\uparrow\downarrow-\uparrow
\downarrow\uparrow-\downarrow\uparrow\uparrow)+udu(2\uparrow\downarrow
\uparrow-\uparrow\uparrow\downarrow-\downarrow\uparrow\uparrow)
+duu(2\downarrow\uparrow\uparrow-\uparrow\uparrow\downarrow-\uparrow
\downarrow\uparrow)]$,
$\psi_{n\uparrow}  = -\frac{1}{3\sqrt{2}}[ddu
(2\uparrow\uparrow\downarrow-\uparrow
\downarrow\uparrow-\downarrow\uparrow\uparrow)
+dud(2\uparrow\downarrow\uparrow-\uparrow\uparrow\downarrow
- \downarrow\uparrow\uparrow)+udd(2\downarrow\uparrow\uparrow
-\uparrow\uparrow\downarrow-\uparrow\downarrow\uparrow)]$,
$\psi_{\Lambda^{o}\uparrow}  = \frac{1}{2}(uds-dus)(\uparrow\downarrow\uparrow
-\downarrow\uparrow\uparrow)$,
$\psi_{\Sigma^{+}\uparrow}  = \frac{1}{\sqrt{6}}uus
(2\uparrow\uparrow\downarrow
-\uparrow\downarrow\uparrow-\downarrow\uparrow\uparrow)$,
$\psi_{\Sigma^{-}\uparrow}  = \frac{1}{\sqrt{6}}dds
(2\uparrow\uparrow\downarrow
-\uparrow\downarrow\uparrow-\downarrow\uparrow\uparrow)$,
$\psi_{\Xi^{o}\uparrow}  = \frac{1}{\sqrt{6}}ssu(2\uparrow\uparrow\downarrow
-\uparrow\downarrow\uparrow-\downarrow\uparrow\uparrow)$,
$\psi_{\Xi^{-}\uparrow}  = \frac{1}{\sqrt{6}}ssd(2\uparrow\uparrow\downarrow
-\uparrow\downarrow\uparrow-\downarrow\uparrow\uparrow)$,
$\psi_{\Sigma^{o}\uparrow}  = \frac{1}{2\sqrt{3}}(uds+dus)
(2\uparrow\uparrow\downarrow
-\uparrow\downarrow\uparrow-\downarrow\uparrow\uparrow)$,
$\psi_{\Delta^{++}\uparrow}  =
uuu\uparrow\uparrow\uparrow$,
$\psi_{\Delta^{+}\uparrow}  =
\frac{1}{\sqrt{3}}(uud+udu+duu)\uparrow\uparrow\uparrow$,
$\psi_{\Delta^{o}\uparrow}  =
\frac{1}{\sqrt{3}}(udd+dud+ddu)\uparrow\uparrow\uparrow$,
$\psi_{\Delta^{-}\uparrow}  =
ddd\uparrow\uparrow\uparrow$,
$\psi_{\Sigma^{*+}\uparrow}  =
uus \uparrow\uparrow\uparrow$,
$\psi_{\Sigma^{*o}\uparrow}  =
\frac{1}{\sqrt{2}}(uds+dus)\uparrow\uparrow\uparrow$,
$\psi_{\Sigma^{*-}\uparrow}  =
dds\uparrow\uparrow\uparrow$,
$\psi_{\Xi^{*o}\uparrow}  =
uss\uparrow\uparrow\uparrow$,
$\psi_{\Xi^{*-}\uparrow}  =
dss\uparrow\uparrow\uparrow$,
$\psi_{\Omega^{-}\uparrow}  =
 sss\uparrow\uparrow\uparrow$.

\end{document}